\documentclass{article}
\usepackage{graphicx} 
\usepackage{array}
\usepackage{amsmath}
\usepackage{hyperref}
\usepackage{bbm}
\newcolumntype{x}[1]{>{\centering\let\newline\\\arraybackslash\hspace{0pt}}p{#1}}
\usepackage[round]{natbib}   
\usepackage[linguistics,edges]{forest}
\usepackage{geometry}

\title{  \large A non-parametric approach for estimating the correlation between log-rank test statistics with applications to a conjunctive power calculation}

\author{Anne Lyngholm Sørensen, Paul Blanche, Henrik Ravn, \\ and Christian Pipper}

\begin{document}
\maketitle
\section*{Abstract}

We present a method for estimating the correlation between log-rank test statistics evaluating separate null hypotheses for two time-to-event endpoints. The correlation is estimated using subject-level data by a non-parametric approach based on the independent and identically distributed (iid) decomposition of the log-rank test statistic under any alternative. Using the iid decomposition, we are able to make an assumption-lean estimation of the correlation. A motivating example using the developed approach is provided. Here, we illustrate how the suggested approach can be used to give a realistic quantification of expected conjunctive power that can guide the design of a new randomized clinical trial using historical data. Finally, we investigate the method's finite sample properties via a simulation study that confirms unbiased and consistent behavior of the proposed approach. In addition, the simulation study gives insight into the effects of censoring on the correlation between the log-rank test statistics.

\section{Introduction}
The majority of clinical trials have more than one endpoint of interest. For multi-endpoint trials, a formal hierarchy among confirmatory endpoints is often defined to reflect their relative importance and to control type-I-error. One frequent example concerns co-primary endpoints. Here, a trial is successful if all primary endpoints demonstrate treatment efficacy. Another frequently encountered example is the use of a single primary and multiple confirmatory secondary endpoints. Here, endpoints are ordered in a strict hierarchy, where a successful trial is one that demonstrates efficacy on the primary endpoint upon which one can proceed to assess efficacy on confirmatory secondary endpoints. The motivating example of this paper focuses on the latter situation. Specifically, we exemplify the developed methodology in the context of multiple time-to-event endpoints that are arranged in a hierarchy of one primary and multiple confirmatory secondary endpoints and where treatment efficacy is assessed via a gate-keeping procedure or ''Fixed sequence method´´ and to control overall type-I-error \citep{Bauer1998,Westfall2001, fda2022multiple}. 

During the design phase of a trial with confirmatory secondary endpoints, an important objective is to optimize both the expected power to demonstrate efficacy on the primary endpoint and the expected power to demonstrate efficacy on both primary and all confirmatory secondary endpoints.  We use the term conjunctive power to denote the power of rejecting all hypotheses of interest simultaneously \citep{Senn2007}. It is well known, that the correlation between the test statistics affects the conjunctive power; a higher correlation between the tests increases the conjunctive power \citep{Senn2007}. Hence, to adequately assess conjunctive power, we need a reliable estimate of the correlation between the involved test statistics. 

This paper is concerned with time-to-event data and the comparison of hazard rates between treatment groups using a log-rank test. Correlation is challenging to summarize between time-to-event endpoints and consequently also test statistics \cite[chap 4]{Hougaard2000}. A challenge in the context of time-to-event studies is censoring resulting from limited follow-up time, which will complicate estimation procedures. Despite challenges, several advances have been made in estimating the correlation between time-to-event endpoints and time-to-event test statistics. In oncology trials, several research papers have focused on estimating the correlation between time to progression-free survival and overall survival. See \cite{Meller2019} for an overview. Many of these methods rely on parametric modeling of the joint distribution of event times, for instance via copulas. The method proposed in \cite{Meller2019} and later reused in \cite{danzer2025} models the joint endpoint distribution based on an illness-death model. A specific example of directly using a copula for parametric modeling of the correlation between time-to-event endpoints is the iterative multiple imputation method from \citep{Schemper13}. Here, a Gaussian copula is assumed and the correlation is estimated using Monte Carlo simulations. Another example where copulas have been utilized for estimating the correlation between log-rank test statistics is \citep{Sugimoto2013}. Similarly to our work, the proposed method was motivated by power and sample size calculations based on historical data. A limitation of copula models and other parametric approaches to model the joint behavior of the subject-level event times is the assumption of a specific dependence structure. This means that for the estimates of, e.g., the correlation to be correctly estimated, the copula model of choice must be correctly specified.

In this paper, we propose a non-parametric method for estimating the correlation between log-rank test statistics. In particular, the method does not require assumptions about the joint behavior of subject-level data to produce unbiased correlation estimates. Our proposal enables us to bypass modeling assumptions on the time-to-event endpoints entirely and directly assess the simultaneous behavior of the log-rank test statistics. To achieve this, we identify the influence function of the log-rank test statistic under any alternative by decomposing it into sums of iid random variables. The influence functions can then be used to characterize the asymptotic joint distribution of the test statistics as a particular multivariate normal distribution \citep{Pipper2012}. We use this characterization to estimate the between test statistic correlations using subject-level data such as data from a historical trial. The resulting method is also computationally efficient as it does not require simulation or bootstrapping of subject-level data to estimate the correlation. We illustrate our approach through a motivating example, where we use it to evaluate potential testing hierarchies of an upcoming trial based on expected conjunctive power using correlations estimated from a similar historical trial. The results provide valuable insights that may guide decisions about the hierarchy. Lastly, we show that the proposed method provides unbiased and consistent estimates through a simulation study. 

The paper is structured as follows. Section 2 will introduce the motivating example of calculating conjunctive power using four endpoints' marginal powers and the between-test-statistic correlations. Our proposed method for estimating the correlation between log-rank test statistics is presented in Section 3. This section will introduce the log-rank test statistic in a counting process setup before decomposing it. Furthermore, it will give an expression of the correlation estimator. The suggested approach is then applied to the motivating example in Section 4. Section 5 contains the simulation study with results, showing that the estimator is unbiased and consistent where we further note that the computation time is fast. The paper concludes with a discussion of the proposed method including the limitations of the method and planned future work in Section 6. 

\section{Motivation: Conjunctive power} \label{sec:conj_power}

Consider an upcoming RCT, trial $X$, which compares a new treatment to control. The trial is planned to include one confirmatory primary endpoint and three confirmatory secondary endpoints. The primary endpoint and secondary endpoints are time-to-event and the trial will use a hierarchical testing procedure to test the four hypotheses. This means that the hypotheses in the trial are tested at the type-I-error level, $\alpha$, in a prespecified order and the family-wise error rate is preserved. The scheme starts with testing the first/primary null hypothesis in the hierarchy. If this null is rejected at level $\alpha$, then the second test in the order is tested at level $\alpha$. Continuation to the third test is allowed when the second null hypothesis is rejected, and continuation to the fourth test is allowed when the third null hypothesis is rejected. Hence, the testing stops when a null hypothesis is not rejected or when all null hypotheses have been tested \citep{Bauer1998, fda2022multiple}. Trial $X$ will closely resemble the concluded SELECT trial \citep{SELECT2023} and use the same type of primary endpoint and secondary endpoints. SELECT is a cardiovascular outcomes trial from pharmaceutical company Novo Nordisk investigating the cardiovascular effects of semaglutide in patients with type 2 diabetes compared to placebo. As in SELECT, the primary endpoint will be time-to-first-occurrence of 3-point MACE (major adverse cardiovascular events), consisting of cardiovascular death (CVD), non-fatal stroke and non-fatal myocardial infarction (MI). Confirmatory secondary endpoints are CVD, all-cause death (ACD) and a composite of heart failure hospitalization and CVD which we abbreviate to HFC. We will use the data from SELECT to guide some decisions concerning the design of trial $X$.

In trial $X$, the hierarchical testing procedure will start with the primary endpoint, MACE. The order of the secondary endpoints is undecided, and we would like to suggest an ordering by investigating which ordering would maximize the expected number of rejected null hypotheses. For that purpose, we will estimate the conjunctive power of all subsets of the tests of the four endpoints. Here, a higher conjunctive power translates to a higher probability of rejecting the null hypotheses. We will now define the probability of rejecting multiple null hypotheses simultaneously in a general sense for our four endpoints, the formulation is not specific to the log-rank test but could be for any standardized test statistic. 

We denote the four null hypotheses by $H_0^{MACE}, H_0^{CVD}, H_0^{ACD},$ and $ H_0^{HFC}$. Let $R^j \in  \{ 0,1\}$ be an indicator of whether $H_0^j$ is rejected in favor of the alternative $H_A^j$ with $j \in \{MACE, CVD, ACD, HFC\}$. With this notation, the conjunctive power of rejecting all four null hypotheses is given as the probability: 
\begin{align*}
P \left(  \{R^{MACE}=1\} \cap \{R^{CVD}=1\} \cap \{R^{ACD}=1\} \cap \{R^{HFC}=1\}  \right).    
\end{align*}
Note that this formulation is not dependent on a specific testing order. It is the probability of all null hypotheses being rejected regardless of the specific testing order. Thus it is equivalent to the probability of rejecting all four null hypotheses when the endpoints are co-primary. We will add the element of a specific order later. With the formal notation in place, we can now illustrate how we calculate the conjunctive power. For the scenario of rejecting all four null hypotheses, we let $\mathbf{Z} = (Z^{MACE}, Z^{CVD}, Z^{ACD}, Z^{HFC})^T$ be a vector of standardized test statistics. Note that $\mathbf{Z}$ can be any set of standardized test statistics. We expect that asymptotically $\mathbf{Z}$ follows a multivariate normal distribution, that is:
\begin{align*} 
\mathbf{Z} =\begin{pmatrix}Z^{MACE}\\
Z^{CVD}\\
Z^{ACD} \\
Z^{HFC}
\end{pmatrix} &\sim  N
\begin{bmatrix}
\begin{pmatrix}
\delta^{MACE}\\
\delta^{CVD}\\
\delta^{ACD} \\
\delta^{HFC}
\end{pmatrix}\!\!,&
\begin{pmatrix}
1 & \rho^{MACE,CVD} & \rho^{MACE,ACD} & \rho^{MACE,HFC} \\
 & 1 & \rho^{CVD, ACD} & \rho^{CVD, HFC} \\
 &  & 1 & \rho^{ACD, HFC} \\
&  &  & 1
\end{pmatrix}
\end{bmatrix}.
\end{align*}
Here $\rho^{j,k}$ is the correlation between the test statistics for the endpoints $j$ and $k$ and $\delta^j$ is the mean of the non-centrality parameter of the $j$th test statistic. For each endpoint $j$, we have $\delta^j = \mathbf{E}(Z^j) = z_{\alpha}+z_{\beta_j}$, where $z_x$ is the $x$'th quantile of the standard normal distribution. Thus, $\delta^j$ is the expected $z$-score needed to achieve power $1-\beta_j$ with significance level $\alpha$ for a one-sided test \citep{proschan2021}. The power $1-\beta_j$ is specific to the outcome's number of events. For a trial with a primary endpoint tested at a significance level of 2.5\% and at 90\% power, the expected z-score is $\delta^{j} = z_{0.025}+z_{0.1} = 3.24$ under an alternative. 

From the distribution of $\mathbf{Z}$, we can approximate the conjunctive power of rejecting all four hypotheses by $P_{\mathbf{Z}} \left( \cap_{j} \{R^j=1\} \right)=P_{\mathbf{Z}} \left( \cap_{j} \{Z^j\in B^{j}\} \right)$, where $B^{j}$ denotes the rejection region of $Z^{j}$. Given $\delta^j$ and $\rho^{j,k}$, one can optimize the order of the tests of the confirmatory secondary endpoints in a three-step procedure such as:
\begin{align*}
   \text{step 1: }&  \operatorname*{argmax}_x  P_{\mathbf{Z}} \left( \cap_{j \in \{MACE,x\}} \{R^j=1\} \bigm| x \in \{CVD, ACD, HFC\} \right) \\
   \text{step 2: }& \operatorname*{argmax}_y  P_{\mathbf{Z}} \left( \cap_{j \in \{MACE,x,y\}} \{R^{j}=1\} \bigm| y \in \{CVD, ACD, HFC\} \setminus \{ x \} \right) \\
   \text{step 3: } & \text{add the remaining endpoint } w \in \{CVD, ACD, HFC\} \setminus \{ x, y \}  
\end{align*}

The above scheme prioritizes maximizing the number of rejected null hypotheses; other schemes might be more relevant, but are not considered here, as they would typically be based on clinical relevance or similar, which is unrelated to the correlation.

\subsection{Estimation of conjunctive power of trial $X$} \label{sec:trial_X_1}

We want to use the above scheme to decide the ordering of the tests of confirmatory secondary endpoints in trial $X$. We wish to use the ordering that maximizes the expected number of rejected null hypotheses. To estimate conjunctive power, we need estimates of the expected test scores $\delta^j$ and the expected correlations between $\rho^{j,k}$. 

The trial will be designed to have a power of 90\% of rejecting the null hypotheses of the primary endpoint (MACE) under the alternative at a significance level of 2.5\%. Thus, the expected test score for the primary endpoint is $\delta^P = z_{0.025} + z_{0.1} = 3.24$. As we are interested in hazard rates, we can write up the relationship between $\delta^P$ and the hazard ratio ($HR$). For a $HR=0.83$, we have since $\log(HR)\sqrt{d/4} \approx 3.24$, that $d = 1211$ which is the number of events needed for a power of 90\% \citep{proschan2021}. Suppose, that we expect that the hazard ratios and number of events will be closely resembling those observed in the SELECT trial \citep{SELECT2023}. Then we will have the following information available, as seen in Table \ref{tab:select_secondary}. Here we have added each endpoint's marginal power, calculated as $\Phi(\delta-1.96)$, where $\Phi$ is the cumulative distribution function of the standard normal distribution. In the table, we have also added how many of the events were CVDs since all endpoints, primary and secondary, contain CVD. The sharing of events will most likely lead to a high correlation between the endpoints and their test statistics.

\begin{table}[h]
    \centering
    \begin{tabular}{l|ccccc}
         Endpoints & $HR$ & Events & CVD events & $\delta$ & Marginal power \\
         \hline
         MACE & 0.83 &  1211 & 410 & 3.24 & 90\% \\
         Cardiovascular death (CVD) & 0.85 &  485 & 485 & 1.79 & 43\% \\
         All-cause death (ACD) & 0.80 & 830 & 485 & 3.21 & 90 \% \\
         HF comp. (HFC) & 0.80 & 660 & 445 & 2.87 & 82 \% \\
    \end{tabular}
    \caption{Hazard ratios, events, estimated test scores $\delta$ and the corresponding marginal power. Hazard ratios and event counts are estimated from observed data from SELECT.}
        \label{tab:select_secondary}
\end{table}

If conjunctive power only depended on marginal power, the optimal choice for the second endpoint in the hierarchy would be ACD as this endpoint has the highest marginal power of the three. However we know that conjunctive power is dependent on the correlations between the test statistics, thus to inform a hierarchical testing order based on conjunctive power, we need the input described in Table \ref{tab:conj_power}. 
    \begin{table}[h]
        \centering
        \begin{tabular}{l|ccc}
             Secondary endpoints & Cardiovascular death (CVD) & All-cause death (ACD)  & HF comp. (HFC) \\ \hline
             $\delta = \log(HR)\sqrt{d/4}$ & 1.79 & 3.21 & 2.87    \\
             Correlation w. MACE-3 & ? & ? & ?  \\
             Correlation w. CVD & - & ? & ?   \\
             Correlation w. ACD & - & - & ?   
        \end{tabular}
        \caption{Test scores for the three confirmatory endpoints. To calculate the conjunctive power we need estimates of the correlation between the endpoints. Missing estimates are indicates with a question mark (?).} 
        \label{tab:conj_power}
    \end{table}

We wish to not rely on a multivariate survival distribution for the endpoints or test statistics to estimate the correlation between the test-statistics. Section \ref{sec:est_corr} introduces a novel method to estimate the correlation between two log-rank test statistics. It uses a decomposition of the log-rank test statistic under an alternative, which is used to estimate the correlation non-parametrically.

\section{Estimating the correlation using a decomposition of the log-rank test statistic} \label{sec:est_corr}

\subsection{Setup and notation}

Consider an RCT that compares a new treatment $(A = 1)$, with a control treatment $(A = 0)$ with $n_1$ participants randomly assigned to new treatment, $n_0$ assigned to control treatment, and consequently with a total participant size $n=n_0+n_1$. Let $T_{ij}$ denote the underlying time-to-event since randomization for participant $i$ $(i = 1,\dots,n)$ for endpoint $j \in \{ 1, 2\}$. For ease of exposition, we consider only two endpoints in this section. Results are straightforward to extend to more than two endpoints. 

We consider an event-driven trial, where end-of-trial is when the proportion of primary events reaches a percentage $q$. Let the stopping time $\hat{\xi}$ be the date where $q$ is reached. For a given date of entry $U_i$ for subject $i$, the limiting value $\xi$ of $\hat{\xi}$ is characterized as the date of analysis that satisfies $P(U_i + T_{i} \leq \xi)=q$. Right-censoring may occur at some calendar time $D_{i}$ and is enforced at the end of the trial such that the time from randomization to censoring for subject $i$ is defined as $C_{i}(\hat{\xi}) = \hat{\xi}\wedge D_{i}-U_i$, thus we observe the right-censored version of the event times $\overline{T}_{ij}(\hat{\xi}) = T_{ij}\wedge C_{i}(\hat{\xi})$. For now we substitute $\hat{\xi}$ by the unknown but fixed $\xi$. This is done to formally avoid conditioning on the future. We end the section by showing that this is a technicality and that the derived decomposition of the log-rank test statistic is still valid when $\xi$ is substituted by $\hat{\xi}$. 

Let $t \in [0,\tau(\xi)]$ be a time point between study start and study duration $\tau(\xi)$, here $\tau(\xi)$ is the maximum difference between end-of-trial $\xi$ and entry $U_i$, thus $\tau(\xi) = \max(\xi - U_i)$. The counting process $N_{ij}(t, \xi)$, the at-risk process $Y_{ij}(t,\xi)$, and filtration $\mathcal{F}_{ij}(t,\xi)$ for the $j$th event type in the $i$th subject are then defined as:
\begin{align*}
    &N_{ij}(t,\xi)=I(T_{ij}\leq t, \: T_{ij}\leq C_{i}(\xi)),\\
    &Y_{ij}(t,\xi)=I(T_{ij}\geq t,\: C_{i} (\xi)\geq t),\\
    &\mathcal{F}_{ij}(t,\xi)=\sigma\{N_{ij}(s,\xi), Y_{ij}(s,\xi), A_{i}, U_{i}:\: s\leq t\}.
\end{align*}
where $\mathcal{F}_{ij}(t,\xi)$ is the smallest $\sigma$-algebra spanned by $\{N_{ij}(s,\xi),Y_{ij}(s,\xi),A_{i}\}_{0\leq s\leq t}$.

Assuming that entry times and thus censoring is independent of both underlying event times and randomized treatment $A$, the counting process intensity with respect to the filtration $\mathcal{F}_{ij}$ is then:
\begin{align*}
\lambda_{ij}(t,\xi)=Y_{ij}(t,\xi)\{\gamma_{0j}(t)(1-A_{i})+\gamma_{1j}(t)A_{i}\},
\end{align*}
where $\gamma_{Aj}(t)$ denote the hazard rate at time $t$ for treatment arm $A$, event type $j$. With the above notation the log-rank test statistic for the $j$th endpoint is defined as:
\begin{align} \label{eq:logrank_iid}
G_{j}(\xi)=\frac{1}{n}\sum_{i=1}^{n}\int_{0}^{\tau(\xi)}\{A_{i}-E_{j}(s,\xi)\}N_{ij}(ds,\xi), \quad \text{with,} \quad E_{j}(t,\xi)=\frac{\sum_{i=1}^{n}A_{i}Y_{ij}(t,\xi)}{\sum_{i=1}^{n}Y_{ij}(t,\xi)}.
\end{align}
Here $G_{j}(\xi)$ is the numerator of the log-rank test statistic for event type $j$, $E_{j}(t,\xi)$ is the proportion of participants assigned to $A = 1$ at risk for event $j$ at time $t$ and $N_{ij}(ds,\xi) = N_{ij}(s,\xi) -\lim_{h\searrow 0} N_{ij}(s-h,\xi)$. For later use we also note that, $E_{j}(t,\xi)$ has the following uniform limit in probability: 
\begin{align*}
e_{j}(t,\xi)=\frac{P(T_{ij}\geq t, C_{i}(\xi)\geq t,\: A_{i}=1)}{P(T_{ij}\geq t, C_{i}(\xi)\geq t)}.
\end{align*} 

The log-rank test is used to test the hypothesis of no difference between two cumulative hazards. We define the null hypotheses of interest as $H_{0j}:\:\: \gamma_{0j}(t)=\gamma_{1j}(t)$ for all $t$ versus the alternative $H_{Aj}:\:\: \gamma_{0j}(t) \neq \gamma_{1j}(t)$ for some $t$.

\subsection{Decomposition}
Our correlation estimate is based on a decomposition that, when properly scaled and centered, approximates $G_{j}(\xi)$ by the sum  of the so-called influence functions $\Phi_{ij}$ which are zero mean iid random variables \citep{tsiatis2006}. Specifically, we show that: 
\begin{align*}
\sqrt{n}\{G_{j}(\xi)-\mathbf{E}(G_{j}(\xi))\}=\frac{1}{\sqrt{n}}\sum_{i=1}^{n}\Phi_{ij}(\xi)+o_{P}(1).  
\end{align*}
In Appendix \ref{app:inf} we derive $\Phi_{ij}(\xi)$ as:
\begin{align}  \label{eq:phi}
\Phi_{ij}(\xi)=&[A_{i}-e_{j}(\overline{T}_{ij}, \xi)]I(T_{ij} \leq C_{i}(\xi)) -\int_{0}^{\tau(\xi)} [A_{i}-e_{j}(t,\xi)] Y_{ij}(t,\xi)\gamma_{0j}(t) dt \\ \nonumber
&-\int_{0}^{\tau(\xi)}Y_{ij}(t,\xi) A_i [1-e_{j}(t,\xi)][\gamma_{1j}(t) - \gamma_{0j}(t)]dt \\ \nonumber
&+\int_{0}^{\tau(\xi)}[A_{i}Y_{ij}(t,\xi)-P(T_{ij}\geq t,\: C_{i}(\xi)\geq t,\: A_{i}=1)][1-e_{j}(t,\xi)]^{2}[\gamma_{1j}(t)-\gamma_{0j}(t)]dt\\ \nonumber
&+\int_{0}^{\tau(\xi)}[(1-A_{i})Y_{ij}(t,\xi)-P(T_{ij}\geq t,\: C_{i}(\xi)\geq t,\: A_{i}=0)][e_{j}(t,\xi)]^{2}[\gamma_{1j}(t)-\gamma_{0j}(t)]dt.\nonumber 
\end{align} 
For estimating the between test statistic correlation  we  use the standardized version of the influence function. With $P(A = 1)=\pi(A)$ the standardized version of the influence function is given by:
\begin{align} \label{eq:denominf}
\Phi_{ij}^*(\xi)= \frac{1}{\sqrt{\pi_A(1-\pi_A)P(T_{ij} \leq \tau(\xi))}} \times \Phi_{ij}(\xi).
\end{align}
By stacking the iid decomposition \citep{Pipper2012}, we can estimate the correlation between  the log-rank test statistics as:
\begin{align}\label{eq:coreq}
corr(G_{1}(\xi),G_{2}(\xi))=\rho_{1,2}(\xi)=\frac{\mathbf{E}(\Phi_{i1}^*(\xi)\Phi_{i2}^*(\xi))}{\sqrt{\mathbf{E}\big[\{\Phi_{i1}^{*}(\xi)\}^2\big]\mathbf{E}\big[\{\Phi_{i2}^{*}(\xi)\}^2\big]}}.
\end{align}

As a final step we relax the assumption of a known date of stopping; $\hat{\xi}$. Instead, we will assume that $\hat{\xi}\overset{P}{\rightarrow}\xi$ and consider the log-rank test as a process in $\xi$, that is:
\begin{align*}
    \mathcal{W}_n(\xi) = \sqrt{n}\{ G_j(\xi) - \mathbf{E}(G_j(\xi)) \}.
\end{align*}
Based on  \cite[chap 19]{vaart_1998_asymp} we argue that the above process is tight as a process of $\xi$. Consequently,
\begin{align*}
\mathcal{W}_n(\hat{\xi}) - \mathcal{W}_n(\xi)=o_p(1).
\end{align*}
We conclude that the decomposition remains valid when we substitute $\hat{\xi}$ for $\xi$.

\subsection{Plug-in estimation} \label{sec:plugin}

To estimate (\ref{eq:coreq}) in practice, we will use plug-in estimates. The formula for the estimated correlation is:
\begin{align*}
\hat{\rho}^{j,k}(\hat{\xi})=\frac{1}{n-1}\frac{\sum_{i=1}^n (\hat{\Phi}^*_{ij}(\hat{\xi})-\overline{\Phi}_{ij}(\hat{\xi}))(\hat{\Phi}^*_{ik}(\hat{\xi})-\overline\Phi_{ik}(\hat{\xi}))}{\sqrt{\frac{1}{n-1}\sum_{i=1}^n(\hat{\Phi}^*_{ij}(\hat{\xi})-\overline{\Phi}_{ij}(\hat{\xi}))^2}\sqrt{\frac{1}{n-1}\sum_{i=1}^n(\hat{\Phi}^*_{ik}(\hat{\xi})-\overline{\Phi}_{ik}(\hat{\xi}))^2}}.
\end{align*}
Where $\overline{\Phi}_{ij}(\hat{\xi}) = \frac{1}{n} \sum_{i=1}^n \hat{\Phi}^*_{ij}(\hat{\xi})\approx 0$.

Let $\hat{e}(t, \hat{\xi})=\frac{\sum_i Y_{ij}(t,\hat{\xi})A_i}{\sum_i Y_{ij}(t,\hat{\xi})}$. For the hazards, we use $\hat{\gamma}_{Aj}(t)dt = d\hat{\Gamma}_{Aj}(t)$, where $\hat{\Gamma}_{Aj}(t)$ is the Nelson-Aalen estimator of the cumulative hazard function at time $t$ for treatment $A$ endpoint $j$. Thus, $\hat{\Gamma}_{Aj}(t)$ is $\sum_{t} \frac{d_{Aj}(t)}{Y_{.Aj}(t)}$ and $\gamma_{Aj} (t) dt = \frac{d_{Aj}(t)}{Y_{.Aj}(t)}$ for all $t$ where $d_{Aj}(t)$ is the number of failures at $t$ in treatment arm $A$ for endpoint $j$, and $Y_{.Aj(t)}$ is the risk set for treatment $A$ at $t$ for endpoint $j$. Lastly, the two probabilities, $P(T_{ij}\geq t,\: C_{i}(\hat{\xi})\geq t,\: A_{i}=1)$ and $P(T_{ij}\geq t,\: C_{i}(\hat{\xi})\geq t,\: A_{i}=0)$ are the probability of being at risk in each arm and are estimated using $\frac{\sum_i Y_{ij}(t,\hat{\xi})A_i}{n}$ and $\frac{\sum_i Y_{ij}(t,\hat{\xi})(1-A_i)}{n}$. With $\hat{\pi}_A =\frac{\sum_{i=1}^n A_i}{n}$ being the observed probability of allocation to treatment $A=1$, the full expression of $\hat{\Phi}^*_{ij}$ is:
\begin{align*}
    \hat{\Phi}^*_{ij} =& \frac{1}{\sqrt{\hat{\pi}_A\left(1-\hat{\pi}_A \right)\frac{\sum_{t=0}^{\tau(\hat{\xi})}d_{0j}(t)+d_{1j}(t)}{n}}} \times \\
    &\Bigg( \Bigg[ A_i -  \frac{\sum_{i=1}^{n}A_{i}Y_{ij}(t,\hat{\xi})}{\sum_{i=1}^{n}Y_{ij}(t,\hat{\xi})} \Bigg]I(T_{ij}\leq C_i(\hat{\xi}))-A_i\sum_{t=0}^{\overline{T}_{ij}(\hat{\xi})}\frac{d_{0j}(t)}{Y_{.0j}(t, \hat{\xi})}+\sum_{t=0}^{\overline{T}_{ij}(\hat{\xi})}\hat{e}(t,\hat{\xi})\frac{d_{0j}(t)}{Y_{.0j}(t, \hat{\xi})} \\
    &- A_i\sum_{t=0}^{\overline{T}_{ij}(\hat{\xi})} [1 - \hat{e}(t,\hat{\xi})]\Bigg[ \frac{d_{1j}(t)}{Y_{.1j}(t, \hat{\xi})} - \frac{d_{0j}(t)}{Y_{.0j}(t, \hat{\xi})} \Bigg] \\
    &+ A_i\sum_{t=0}^{\overline{T}_{ij}(\hat{\xi})} [1 - \hat{e}(t,\hat{\xi})]^2\Bigg[ \frac{d_{1j}(t)}{Y_{.1j}(t, \hat{\xi})} - \frac{d_{0j}(t)}{Y_{.0j}(t, \hat{\xi})} \Bigg]  \\
    &- \sum_{t=0}^{\overline{T}_{ij}(\hat{\xi})} \frac{Y_{.1j}(t, \hat{\xi})}{n} [1 - \hat{e}(t,\hat{\xi})]^2\Bigg[ \frac{d_{1j}(t)}{Y_{.1j}(t, \hat{\xi})} - \frac{d_{0j}(t)}{Y_{.0j}(t, \hat{\xi})} \Bigg] \\
    &+(1-A_i)\sum_{t=0}^{\overline{T}_{ij}(\hat{\xi})} \hat{e}(t,\hat{\xi})^2\Bigg[ \frac{d_{1j}(t)}{Y_{.1j}(t, \hat{\xi})} - \frac{d_{0j}(t)}{Y_{.0j}(t, \hat{\xi})} \Bigg] \\
    &- \sum_{t=0}^{\overline{T}_{ij}(\hat{\xi})} \frac{Y_{.0j}(t, \hat{\xi})}{n} \hat{e}(t,\hat{\xi})^2\Bigg[ \frac{d_{1j}(t)}{Y_{.1j}(t, \hat{\xi})} - \frac{d_{0j}(t)}{Y_{.0j}(t, \hat{\xi})} \Bigg] \Bigg).
\end{align*}

\section{Optimizing the step-wise conjunctive power}

We can now return to our motivating example of designing a new trial using conjunctive power. We want to estimate the correlation between four test statistics using data from the SELECT trial. We can update Table \ref{tab:conj_power} with estimates of the correlation using our suggested estimator (\ref{eq:coreq}) via the plug-in estimator presented in subsection \ref{sec:plugin}. See Table \ref{tab:conj_power_upd} for an updated version of Table \ref{tab:conj_power} with the estimates of the correlations. All endpoints are quite correlated, which is natural as the outcomes share events; CVD is contained in the primary endpoint (MACE), in the HFC outcome, and in the ACD outcome.

    \begin{table}[h]
        \centering
        \begin{tabular}{l|ccc}
             Secondary endpoints & CVD & ACD  & HFC \\ \hline
             $\delta$ & 1.79 & 3.21 & 2.87    \\
             Correlation w. MACE-3 & 0.60  & 0.48  & 0.56  \\
             Correlation w. CVD & - & 0.76  & 0.85   \\
             Correlation w. ACD & - & - & 0.67  
        \end{tabular}
        \caption{Updated table with estimates of the correlations using the suggested estimator.}
        \label{tab:conj_power_upd}
    \end{table}

With the estimates from Table \ref{tab:conj_power_upd} and an assumed power of 90\% to reject the primary null hypothesis at an $\alpha$ level of 2.5\% (i.e. $\delta^{MACE} = 3.24$), the conjunctive power of rejecting all four null hypotheses is 42\% using the multivariate normal distribution described in Section \ref{sec:conj_power}. To give an understanding of the effect of the correlations on the power, we can calculate the conjunctive power under the assumption of independence, thus setting the correlations $\rho^{j,k}=0$ for all $j \neq k$. This gives a conjunctive power of 28\%. Equivalently, we can also get the 28\% power by multiplying the marginal powers of the four null hypotheses (i.e. $0.9\times0.43\times0.90\times0.82 \approx 0.28 $). Instead, assuming that they are perfectly dependent such that the correlations $\rho^{j,k}=1$ for all $j \neq k$, we get a conjunctive power of 43\% which equals the minimum marginal power of the four considered endpoints.

\begin{figure}[ht]
\includegraphics[width=1.0\linewidth]{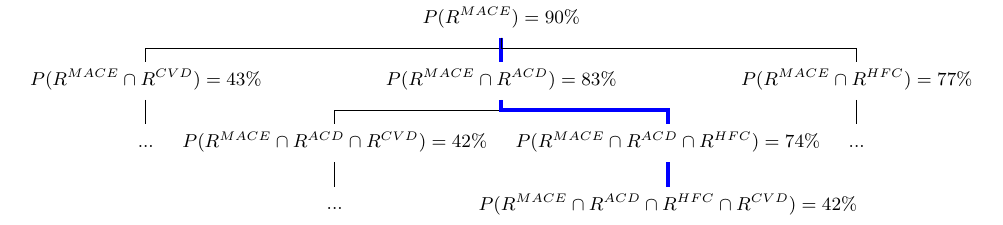}
    \caption{Using conjunctive power to select the ordering of the confirmatory secondary endpoints. The blue line indicates the optimal ordering of the confirmatory secondary endpoints' test with respect to maximizing the conjunctive power.}
    \label{fig:hier_order}
\end{figure}

The conjunctive power can also be calculated for subsets of the null hypotheses. By calculating the power of the different sets, we can use the three-step procedure described in Section \ref{sec:conj_power} to find the testing order that maximizes the conjunctive power at each level of the hierarchy. See Figure \ref{fig:hier_order} where we have illustrated the optimal testing order, where the hierarchy starts with the primary test at the top and the ordering of the secondary endpoints' tests is based on the highest conjunctive power. With the estimates presented in Table \ref{tab:conj_power_upd}, the optimal hierarchy in terms of conjunctive power is $MACE \rightarrow ACD \rightarrow HFC \rightarrow CVD$ for trial $X$ which corresponds to estimated hierarchy-level conjunctive powers of $90\% \rightarrow 83\% \rightarrow74\% \rightarrow 42\%$. This means that the conjunctive power of rejecting the primary and first secondary endpoint is $83\%$ under the alternatives being true.  

In practice, the calculated optimal hierarchy can be used as a statistical perspective for how the endpoints should be arranged. Thus, it will serve more as supporting information than as a strict rule for the design as other considerations also matter. Furthermore, since the calculation is based on historical data, users will need to evaluate how closely they expect the new trial data to resemble the old. Hence, performing several calculations using slightly higher or smaller values of the correlations will often be valuable. We have a more detailed discussion of this topic in Section \ref{sec:disc}.

\section{Simulation study}

\subsection{Simulation aims, set-up and evaluations}

We will via simulations investigate the empirical properties of the estimator with a focus on the accuracy of the method under scenarios with varying sample sizes, data generating distributions, levels of censoring, and correlation. The simulation studies are in some parts inspired by the SELECT trial, but will also cover more general scenarios. The main objective of the simulation study in this paper is to verify that the estimator defined in (\ref{eq:coreq}) is unbiased and consistent. Hence, the focus is on the estimate of the correlation of the log-rank test statistics via our method. 

The simulated data in this section are generated using parametric simulations. We are simulating a trial with 1:1 randomization to either control or treatment where the two endpoints of interest are time-to-event endpoints where we categorize one as a primary endpoint and one as a secondary. We let participants be recruited over 1.5 years uniformly. The trial will be event-driven and will stop once the number of events for the primary endpoints reaches a certain amount specific to the wanted level of censoring. 

The two endpoints will be correlated and we will use different copula models to create correlated survival data. The different copula models are: Gaussian, Clayton, Frank, and Gumbel \citep{Hougaard2000}. For the Gaussian copula we will investigate the correlation under copula parameters of 0, 0.5, and 0.8 which translates to the correlation between the endpoints under no censoring. For the Frank, Clayton, and Gumbel models we choose copula parameters respectively at 4, 1 and 1. This provides correlations of different sizes given the specific simulation setup as presented in the results. We choose to look at different copula models to investigate different correlation structures. From the uniform data generated from the copulas, we transform the data to an exponential distribution. The rate in the treatment arm depends on a given hazard ratio. The data will be exponentially distributed in two forms; constant rates and piecewise constant rates. Two levels of censoring are considered at 80\% and 93\%. Some of the simulation parameters are inspired by the SELECT trial including the censoring level of 93\%. Others are the hazard rates, simulated data size $n_{obs}$ (such as $n_{obs}=17600$), and the hazard ratio. However, most of these simulation parameters also include other values to explore more general scenarios. 

Each scenario of interest is simulated 10,000 times. Depending on the focus of the simulation study, different combinations of the simulation parameters as found in Table \ref{tab:sim_param}, will be used. Also inspired by the SELECT trial, correlation will be introduced by the endpoints being distributionally correlated (via a copula model) but also by letting an endpoint be a composite endpoint containing the events of the other endpoint. We define a composite endpoint in the following manner. Let again $\overline{T}_{ij}$ be the potentially right-censored event time for subject $i$ endpoint $j$. The time of the composite endpoint is defined as $\overline{V}_{i1} = \min \left( \overline{T}_{i1}, \overline{T}_{i2}  \right)$. For simulations with composite endpoints, the correlation will be based on $\overline{V}_{i1}$ and $\overline{T}_{i2}$. An example of a composite endpoint is MACE-3 as described in Section \ref{sec:trial_X_1}. MACE-3 consists of nonfatal myocardial infarction (MI), nonfatal stroke, and cardiovascular death (CVD), thus with CVD being a secondary endpoint in the trial, MACE-3 is a composite of CVD and non-fatal MI and stroke. A list of all simulation parameters is available in Table \ref{tab:sim_param}. We use these endpoints for notation in the simulation study which is also shown in Table \ref{tab:sim_param}.

\begin{table}[ht]
\centering
\begin{tabular}{lp{9cm}}
  \hline
Simulation parameters & Values \\ 
  \hline
    $n_{sim}$ & 10,000 \\ 
  $n_{obs}$ & 400, 4000, 15000, 17600, 40000 \\
  - per arm & 200, 2000, 7500, 8800, 20000 \\ 
Copula model & Gaussian, Clayton, Frank and Gumbel \\ 
- $\theta$ (copula param.) & Gaussian: 0, 0.5, 0.8 \\
& Clayton: 1 \\
& Frank: 4 \\
& Gumbel: 1 \\
Hazard rate & exponential constant and piecewise constant \\
- rates & constant: \\
& $\gamma^{plcb}_{mi,stroke} = 0.017$  and $\gamma^{plcb}_{cvd} = 0.009$  \\
& piecewise constant: change in hazard at $t = 730.5$ days \\
& $\gamma^{plcb}_{mi,stroke}(0-t) = 0.017$ and $\gamma^{plcb}_{mi,stroke}(t-\infty) = 0.017$ \\
& $\gamma^{plcb}_{cvd}(0-t) = 0.007$ and $\gamma^{plcb}_{cvd}(t-\infty) = 0.012$ \\
  Hazard ratios & 0.8 \\  
  Composite endpoint & yes, no \\ 
  Level of censoring & 80\%, 93\% \\ 
   \hline
\end{tabular}
\caption{Simulation parameters for the simulation studies. $n_{sim}$ is the number of simulations per scenario, $n_{obs}$ is the number of simulated participants per data set and $\gamma^x_y$ is the hazard rate for treatment group $x$ for endpoint $y$. When simulating piecewise constant hazard rates the notation is instead $\gamma^x_y(z-w)$ where $z$ indicates the start time and $w$ indicates the end time of the time period considered. Note that $plcb$ stands for placebo.}
\label{tab:sim_param}
\end{table}

We wish to check if the estimator of the correlation between two log-rank test statistics presented in (\ref{eq:coreq}) is unbiased and consistent. We will investigate the performance of the correlation estimate by its bias and observed $2.5\%$ and $97.5\%$ percentiles. To evaluate the bias, we set the true value of the correlation to the empirical correlation estimate from the simulated log-rank test statistics. Thus, in each simulation we collect the log-rank test $z^{j}_i$ and $z^{k}_i$ for the two endpoints $j$ and $k$ for $i=1,\dots,n_{sim}$. The true correlation is then the Pearson correlation between the simulated log-rank test scores which we denote $\tilde{\rho}$. In each simulation, we also collect the estimated correlation using our iid decomposition defined in (\ref{eq:coreq}) denoted $\hat{\rho}^{iid}_i$ where the mean $\overline{\rho}^{iid} = 1/n_{sim} \cdot\sum_{i=1}^{n_{sim}}\hat{\rho}^{iid}_i$ is reported. The estimate of the bias is then calculated as: $\text{bias} = \overline{\rho}^{iid} - \tilde{\rho}$. We will investigate whether the estimator is consistent using the percentile range to confirm that it shrinks with an increase in $n_{obs}$. 

\subsection{Simulation results}

Before we present the simulation results, we highlight the speed of the suggested approach. Using an example data set (which is also available at {\tt{https://github.com/AnneLyng/cor\_logrank}}, the average computation time of the correlation estimate was 0.02 seconds, where the timing was calculated using the {\tt{microbenchmark package}} \citep{microbenchmark} in {\tt{statistical software R}} \citep{RSoft}. This average computation time includes the time using the {\tt{cor function}}. Thus, compared to e.g. bootstrapping, it will provide results faster, making it more applicable for e.g. simulation studies. As the computation time is relative to the system in which the code is run, readers can run the code locally to investigate the computation time, but as the implementation of the approach is relatively simple, we anticipate the method to run fast for all users. 

We will now show that the approach is unbiased and consistent. Table \ref{tab:simbias} and \ref{tab:simselect} show the simulation results. Table \ref{tab:simbias} looks at scenarios with constant hazard rates, no composite endpoints and a trial arm size of 8800 (inspired from SELECT). We find that the suggested method is unbiased for all scenarios considered indifferent of copula model and censoring. We further find that censoring in the scenarios considered decreases the correlation between the test statistics, but the degree varies on copula model. This is expected as the different copula models concentrate the correlation at different timings. As an example, the Clayton copula is concentrated earlier than the Gumbel copula, which is seen as the censoring has a larger effect (more diluting of the size of the correlation) on the Gumbel copula generated data than the Clayton copula. 

Table \ref{tab:simselect} is inspired by the SELECT trial, where the endpoints are composite, the hazard rates are piecewise constant and the trial arm size is 8800. The level of censoring in SELECT on the primary MACE endpoint was $\sim 93 \%$. Similar to before, we find that the approach is unbiased. We note that having composite endpoints affects the correlation as we estimate higher correlations. 

\begin{table}[ht]
\centering
\begin{tabular}{rclrllx{4cm}}
  \hline
Copula & $\theta$ & censoring & bias & $\tilde{\rho}$ & $\overline{\rho}^{iid}$ (2.5\%, 97.5\% percen.)  \\
\hline
Gaussian & 0 & 80\% & 0.006 &- 0.006 & $<0.001$ (-0.015;0.016)\\
- & - & 93\% & 0.009 & -0.009 & $<0.001$ (-0.015;0.015)   \\
- & 0.5 & 80\% & 0.013 & 0.265 & 0.279 (0.260;0.297) \\
- & - & 93\% & 0.002 & 0.203 & 0.205 (0.178;0.232)   \\
- & 0.8 & 80\% &  0.015 & 0.515 & 0.530 (0.531;0.546) \\
- & - & 93\% &  0.006 & 0.452 & 0.458 (0.429;0.486) \\
Clayton & 1 & 80\% &  -0.024 & 0.473 & 0.449 (0.431;0.467) \\
- & - & 93\% &  -0.024 & 0.477 & 0.454 (0.424;0.483)  \\
Frank & 4 & 80\% &  -0.018 & 0.283 & 0.265 (0.246;0.284)  \\
- & - & 93\% &  -0.022 & 0.133 & 0.111 (0.086;0.137)  \\
Gumbel & 1 & 80\% & -0.019 & 0.365 & 0.346 (0.328;0.364)  \\
 & - & 93\% &  -0.018 & 0.250 & 0.232 (0.203;0.261) \\
   \hline
\end{tabular}
\caption{Simulation results for $n_{obs} = 8800$ under varying copula models with constant exponential hazard rate, hazard ratios of 0.8 and non-composite endpoints. Here $\theta$ is the copula parameter, $\tilde{\rho}$ is the simulated true correlation between log-rank tests and $\overline{\rho}^{iid}$ is the average estimated correlation using the iid decomposition accompanied with the observed 2.5\% and 97.5\% percentiles illustrated in the parentheses. The number of simulations was $n_{sim} = 10,000$.}
\label{tab:simbias}
\end{table}

\begin{table}[ht]
\centering
\begin{tabular}{rclrllx{4cm}}
  \hline
Copula & $\theta$ & censoring & bias & $\tilde{\rho}$ & $\overline{\rho}^{iid}$ (2.5\%, 97.5\% percen.)  \\
\hline
Gaussian & 0 & 93\% & 0.014 & 0.611 &  0.624 (0.602;0.647)   \\
- & 0.8 & 93\% &  0.002 & 0.568 & 0.570 (0.547;0.592) \\
Clayton & 1 & 93\% &  -0.005 & 0.528 & 0.523 (0.499;0.546)  \\
Frank & 4 & 93\% &  -0.008 & 0.594 & 0.585 (0.863;0.882)  \\
Gumbel & 1 & 93\% &  -0.010 & 0.586 & 0.576 (0.826;0.845)  \\
   \hline
\end{tabular}
\caption{Simulation results for $n_{obs} = 8800$ under varying copula models with piecewise-constant exponential hazard rate, hazard ratios of 0.8 and composite endpoints. Here $\theta$ is the copula parameter, $\tilde{\rho}$ is the simulated true correlation between log-rank tests and $\overline{\rho}^{iid}$ is the average estimated correlation using the iid decomposition accompanied with the observed 2.5\% and 97.5\% percentiles illustrated in the parentheses. The number of simulations was $n_{sim} = 10,000$}
\label{tab:simselect}
\end{table}

\begin{figure}
    \centering
    \includegraphics[width=0.9\linewidth]{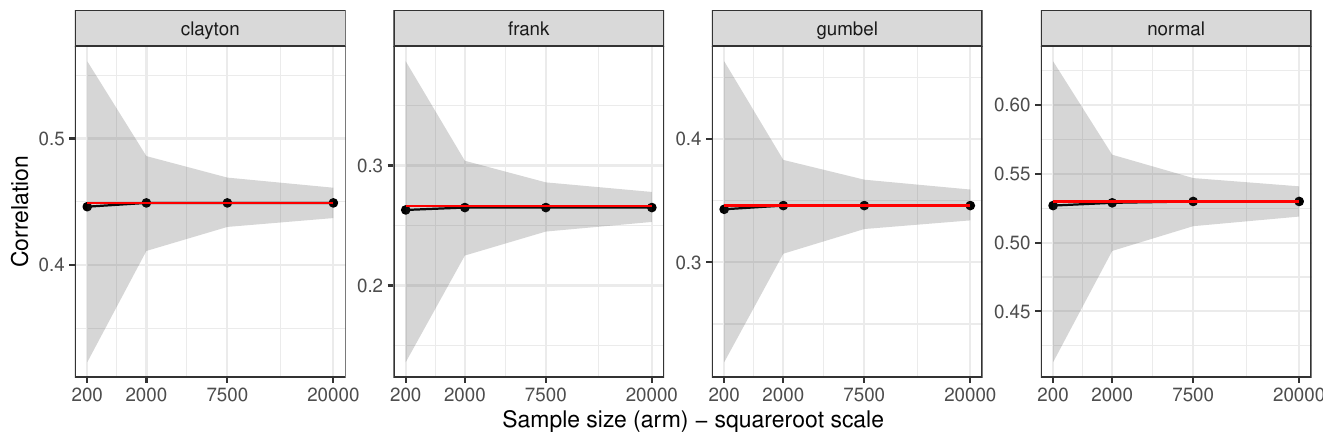}
    \caption{The shaded areas illustrates the 2.5 percentile and 97.5 percentile of the correlation estimates. The red line is the true value and the black line is the mean correlation estimate from the proposed method $\overline{\rho}^{iid}$.}
    \label{fig:consis}
\end{figure}

The variation of the estimate of the correlation is dependent on the sample size, here described by the number of participants per arm. We find, unsurprisingly, that larger sample sizes decrease the variation. Figure \ref{fig:consis} show for the four different copula models how the 2.5 and 97.5 percentile decreases as a function of sample size. Hence, our estimator shows consistent behavior.

\section{Discussion} \label{sec:disc}

Using the iid decomposition of the log-rank test statistic and stacking, we developed a new method to estimate the correlation between log-rank test statistics. The method was applied to a motivating example which showed how it can be used to design new trials by informing the statistically optimal testing order in a hierarchical testing procedure. By simulations, we showed that the method is unbiased and consistent. We will now discuss some of the limitations of the method as applied in this paper which leads to a discussion of planned investigations extending the use of the method.

The use of the correlation between test statistics can be for both designing new trials and but also for analyzing the observed data at the end of the trial. This paper illustrated how it could be used for designing for which we will describe some limitations of. In our motivating example, we applied the method for suggesting a specific arrangement the hierarchy in the design phase of a new trial. The ordering was optimized based on conjunctive power which is dependent on the test statistics expected size and the test statistics' correlations. The correlations are based on a historical trial. Our estimates of conjunctive power used for designing a new trial is at most relevant under the assumption of the new trial's characteristics such as rates and treatment effects mimics the historical trial's. This might be the case for some studies, but is an unrealistic assumption for others. In the case where we expect differences, it is valuable to know how adaptable the method is. While the expected test scores anticipated in the new trials might be set by medical expertise and are easy to vary, adapting the historical data in the correlation calculating to the newly expected rates and treatment effects of new treatments while keeping other characteristics fixed is more complicated. We plan to investigate both how easy (or difficult) it is to adapt the operational characteristics using historical data and what operational characteristics that affect the correlation estimate and to what extend. Thus, we want investigate the characteristics both for a) their adaptability to new trials with expected changes in treatment effect etc. b) understanding what affects the correlation the most. If it is difficult to adapt the historical data to new scenarios, the evaluation of how the correlation changes depending on the operational characteristics, everything else being equal, could serve as another way to quantify uncertainty in e.g. conjunctive power calculations for a future trial.

Another future planned work is using the correlation estimate to optimize the test of a secondary endpoint in a group sequential trial. There are several papers on how to optimize group sequential trials with multiple endpoints in terms of alpha-spending such as \citep{Glimm2010, Tamhane2012, Li2018,danzer2025}. 
Here both \citep{Glimm2010,Li2018} assume a constant correlation. We wish to investigate if the optimal alpha-spending procedures change as the correlation for time-to-event test statistics should not be considered constant. In the paper by \citep{Tamhane2012} there is no such assumption, but their method does not support time-to-event test statistics. Further research is hence needed in this area and can be supported by the method we have proposed. In a recent paper by \citep{danzer2025} both an expression of the covariance matrix of two time-to-event endpoints is provided along with a testing strategy for how to optimize the type-I-error spend for two primary endpoints in an oncology trial setting. Their covariance matrix is derived under an illness-death model exemplified using progression-free survival and overall survival as endpoints. We find that the results presented in \citep{danzer2025} complement the results concerning the correlation estimator in this paper as the method provided in this paper can be used for other endpoints that may not adhere to the illness-death model and provide realistic power calculations. For the planned future research for optimizing the testing of multiple endpoints in a group sequential trial, we wish to investigate optimal designs for not multiple primary endpoints but in scenarios with a single primary endpoint and at least one secondary endpoint. 

In conclusion, our proposed method provides an unbiased and consistent estimate of the correlation without bootstrapping, simulation, or distributional assumptions. Compared to bootstrapping, it is faster to use. Furthermore, if one is interested in the sensitivity to e.g. censoring, it is easy (and fast) to investigate via the method's implementation. See the supplementary information or the repository on GitHub {\tt{https://github.com/AnneLyng/cor\_logrank}} for code and an example script written in {\tt{R}} of how to use the code on simulated trial data.

\noindent {\bf{Conflict of Interest}} \\
\noindent {\it{Anne L. Sørensen, Henrik Ravn and Christian B. Pipper are employees at Novo Nordisk A/S. Henrik Ravn and Christian B. Pipper own stock in Novo Nordisk A/S.}}

\newpage

\appendix

\section*{Appendix} \label{app:inf}

This appendix section will provide a summarized derivation of the influence function $\Phi_{ij}$ which is part of the iid decomposition of the un-standardized log-rank test statistics. For readability we do not emphasize the dependence of the date of stopping $\xi$ in this section as it will not play a visible part in the derivation of the decomposition. 
\begin{align*} 
\sqrt{n}\{G_{j}-\mathbf{E}(G_{j})\}=\frac{1}{\sqrt{n}}\sum_{i=1}^{n}\Phi_{ij}+o_{P}(1).    
\end{align*}
The derivation of $\Phi_{ij}$ can be split into four parts, which we will cover here. The four steps are:
\begin{enumerate}
    \item Use martingale theory to re-formulate $G_{j}$
    \item Work on the compensator part of $G_{j}$
    \item Derive an expression for $\mathbf{E}(G_{j})$
    \item Combining everything
\end{enumerate}

\subsection*{A.1.\enspace Use martingale theory to re-formulate $G_{j}$}
Remember that $G_{j}$ is formulated as:
$$
G_{j}=\frac{1}{n}\sum_{i=1}^{n}\int_{0}^{\tau}\{A_{i}-E_{j}(s)\}dN_{ij}(s),
$$

With $M_{ij}(t)=N_{ij}(t)-\int_{0}^{t}\lambda_{ij}(s)ds$ denoting the counting process martingale, we rewrite $G_{j}$ as:
$$
\sqrt{n}G_{j}=\frac{1}{\sqrt{n}}\sum_{i=1}^{n}\int_{0}^{\tau}\{A_{i}-e_{j}(t)\}dM_{ij}(t)+\frac{1}{\sqrt{n}}\sum_{i=1}^{n}\int_{0}^{\tau}\{A_{i}-E_{j}(t)\}\lambda_{ij}(t)dt+o_{P}(1).
$$
Here the $o_P(1)$ term comes from changing $E_{j}(t)$ to $e_{j}(t)$ in the martingale part of the equation.

\subsection*{A.2.\enspace Compensator part of $G_{j}$} \label{app:sec:comp}

We rewrite the second term $\frac{1}{\sqrt{n}}\sum_{i=1}^{n}\int_{0}^{\tau}\{A_{i}-E_{j}(t)\}\lambda_{ij}(t)dt$ by first using that $\sum_{i=1}^{n} A_i Y_{ij} = E_{j}Y_{.jl}$ and by writing out the expression for $\lambda_{ij}(t)$ and adding 0 twice: ($e_{j}(t)-e_{j}(t)$ and $-nP(T_{ij}\geq t,\: C_{i}\geq t, A_{i}=1)+ nP(T_{ij}\geq t,\: C_{i}\geq t, A_{i}=1)$). We further introduce $Y_{\cdot jl}(t)=\sum_{i=1}^n Y_{ij}(t)$. 
\begin{align}
 & \frac{1}{\sqrt{n}}\sum_{i=1}^{n} \int_{0}^{\tau}\{A_{i}-E_{j}(t)\}\lambda_{ij}(t)dt = \nonumber \\ 
&\frac{1}{\sqrt{n}}\int_{0}^{\tau}[e_{j}(t)-E_{j}(t)] \times \nonumber \\
&[E_{j}(t)Y_{\cdot jl}(t)-nP(T_{ij}\geq t,\: C_{i}\geq t, A_{i}=1)][\gamma_{1j}(t)-\gamma_{0j}(t)]dt \label{eq:comp1} \\
&+\frac{1}{\sqrt{n}}\int_{0}^{\tau}[e_{j}(t)-E_{j}(t)]nP(T_{ij}\geq t,\: C_{i}\geq t, A_{i}=1)[\gamma_{1j}(t)-\gamma_{0j}(t)]dt \label{eq:comp2}\\
&+ \frac{1}{\sqrt{n}}\int_{0}^{\tau}[1-e_{j}(t)][E_{j}(t)Y_{\cdot jl}(t)-nP(T_{ij}\geq t,\: C_{i}\geq t, A_{i}=1)][\gamma_{1j}(t)-\gamma_{0j}(t)]dt \label{eq:comp3}\\
&+\sqrt{n}\int_{0}^{\tau}[1-e_{j}(t)]P(T_{ij}\geq t,\: C_{i}\geq t, A_{i}=1)][\gamma_{1j}(t)-\gamma_{0j}(t)]dt \label{eq:comp4}
\end{align}

We end with four terms in the compensator part of the counting process. Now note that the first term (\ref{eq:comp1}) in the last equality is $o_{P}(1)$ as $[e_{j}(t)-E_{j}(t)] = o_P(1)$ and $[E_{j}(t)Y_{\cdot jl}(t)-nP(T_{ij}\geq t,\: C_{i}\geq t, A_{i}=1)] = O_P(1)$. For the second term (\ref{eq:comp2}) we write out the expressions for $e_{j}(t)$ and $E_{j}(t)$ and use the ``monkey eating its own tail'' trick:
$$
\frac{a}{b} - \frac{a_n}{b_n} = \frac{a b_n - b a_n}{b b_n} = \frac{a b_n - n\cdot ab + n\cdot ab - ba_n}{bb_n}
$$
With $a_n = \sum_{i=1}^{n} A_i Y_{ij}(t), a = P(T_{ij}\geq t,\: C_{i}\geq t, A_{i}=1), b_n = Y_{.jl}(t),$ and $ b = P(T_{ij} \geq t, C_{ij} \geq t)$. Consequently, for the second term (\ref{eq:comp2}):
\begin{eqnarray*}
    &&\frac{1}{\sqrt{n}}\int_{0}^{\tau}[e_{j}(t)-E_{j}(t)]nP(T_{ij}\geq t,\: C_{i}\geq t, A_{i}=1)[\gamma_{1j}(t)-\gamma_{0j}(t)]dt=\\
    &&\frac{1}{\sqrt{n}}\sum_{i=1}^{n}\int_{0}^{\tau}[Y_{ij}(t)-P(T_{ij}\geq t,\: C_{i}\geq t)]e_{j}(t)^{2}[\gamma_{1j}(t)-\gamma_{0j}(t)]dt\\
    &&-\frac{1}{\sqrt{n}}\sum_{i=1}^{n}\int_{0}^{\tau}[A_{i}Y_{ij}(t)-P(T_{ij}\geq t,\: C_{i}\geq t,\: A_{i}=1)]e_{j}(t)[\gamma_{1j}(t)-\gamma_{0j}(t)]dt+o_{P}(1).
\end{eqnarray*}
For the third term (\ref{eq:comp3}) we have 
\begin{eqnarray*}
    &&\frac{1}{\sqrt{n}}\int_{0}^{\tau}[1-e_{j}(t)][E_{j}(t)Y_{\cdot jl}(t)-nP(T_{ij}\geq t,\: C_{i}\geq t, A_{i}=1)][\gamma_{1j}(t)-\gamma_{0j}(t)]dt=\\
    &&\frac{1}{\sqrt{n}}\sum_{i=1}^{n}\int_{0}^{\tau}[1-e_{j}(t)][A_{i}Y_{ij}(t)-P(T_{ij}\geq t,\: C_{i}\geq t, A_{i}=1)][\gamma_{1j}(t)-\gamma_{0j}(t)]dt
\end{eqnarray*}

We know need a definition of $\mathbf{E}(G_{ij})$.

\subsection*{A.3.\enspace Finding an expression for $\mathbf{E}(G_{j})$}
Here it is used that the non-centrality term is approximated as:
$$
\sqrt{n}\mathbf{E}(G_{j})=\sqrt{n}\int_{0}^{\tau}[1-e_{j}(t)]P(T_{ij}\geq t,\: C_{i}\geq t, \: A_{i}=1)[\gamma_{1j}(t)-\gamma_{0j}(t)]dt+o_{P}(1).
$$
This is derived by taking the expecation of $\sqrt{n}G_{j}$:
$$
\sqrt{n}G_{j}=\frac{1}{\sqrt{n}}\sum_{i=1}^{n}\int_{0}^{\tau}\{A_{i}-e_{j}(t)\}dM_{ij}(t)+\frac{1}{\sqrt{n}}\sum_{i=1}^{n}\int_{0}^{\tau}\{A_{i}-E_{j}(t)\}\lambda_{ij}(t)dt+o_{P}(1).
$$
The first part as expectation 0 as it is a martingale. The second part (the compensator part) is shown in Section \ref{app:sec:comp} and reduces to (\ref{eq:comp4}) and a remainder when taking the expectation. 

\subsection*{A.4.\enspace Combining everything}

The fourth term (\ref{eq:comp4}) is seen to be completely deterministic and approximates $\sqrt{n}\mathbf{E}(G_{j})$ which means that in $\Phi_{ij}$ we add and subtract $\sqrt{n}\mathbf{E}(G_{j})$. Thus, we have that when combining all the terms: 
\begin{align*}
\sqrt{n}&\{G_{j}-\mathbf{E}(G_{j})\} = \frac{1}{\sqrt{n}}\sum_{i=1}^{n}\int_{0}^{\tau}\{A_{i}-e_{j}(t)\}dM_{ij}(t) \\
&+\frac{1}{\sqrt{n}}\sum_{i=1}^{n}\int_{0}^{\tau}[Y_{ij}(t)-P(T_{ij}\geq t,\: C_{i}\geq t)]e_{j}(t)^{2}[\gamma_{1j}(t)-\gamma_{0j}(t)]dt\\
    &-\frac{1}{\sqrt{n}}\sum_{i=1}^{n}\int_{0}^{\tau}[A_{i}Y_{ij}(t)-P(T_{ij}\geq t,\: C_{i}\geq t,\: A_{i}=1)]e_{j}(t)[\gamma_{1j}(t)-\gamma_{0j}(t)]dt \\
&+ \frac{1}{\sqrt{n}}\sum_{i=1}^{n}\int_{0}^{\tau}[1-e_{j}(t)][A_{i}Y_{ij}(t)-P(T_{ij}\geq t,\: C_{i}\geq t, A_{i}=1)][\gamma_{1j}(t)-\gamma_{0j}(t)]dt \\
&+\sqrt{n}\int_{0}^{\tau}[1-e_{j}(t)]P(T_{ij}\geq t,\: C_{i}\geq t, A_{i}=1)][\gamma_{1j}(t)-\gamma_{0j}(t)]dt \\
&- \sqrt{n}\int_{0}^{\tau}[1-e_{j}(t)]P(T_{ij}\geq t,\: C_{i}\geq t, \: A_{i}=1)[\gamma_{1j}(t)-\gamma_{0j}(t)]dt \\ &+ o_P(1)
\end{align*}
Where:
\begin{align*}
\Phi_{ij} =& \int_{0}^{\tau}\{A_{i}-e_{j}(t)\}dM_{ij}(t) \\
&+\int_{0}^{\tau}[Y_{ij}(t)-P(T_{ij}\geq t,\: C_{i}\geq t)]e_{j}(t)^{2}[\gamma_{1j}(t)-\gamma_{0j}(t)]dt\\
&-\int_{0}^{\tau}[A_{i}Y_{ij}(t)-P(T_{ij}\geq t,\: C_{i}\geq t,\: A_{i}=1)]e_{j}(t)[\gamma_{1j}(t)-\gamma_{0j}(t)]dt \\
&+ \int_{0}^{\tau}[1-e_{j}(t)][A_{i}Y_{ij}(t)-P(T_{ij}\geq t,\: C_{i}\geq t, A_{i}=1)][\gamma_{1j}(t)-\gamma_{0j}(t)]dt 
\end{align*}

We can simplify this expression by using that $P(T_{ij}\geq t,\: C_{i}\geq t) = P(T_{ij}\geq t,\: C_{i}\geq t, A_{i}=1) + P(T_{ij}\geq t,\: C_{i}\geq t, A_{i}=0)$ and $Y_{ij}(t) = A_iY_{ij}(t) + (1-A_i)Y_{ij}(t)$ in $[Y_{ij}(t)-P(T_{ij}\geq t,\: C_{i}\geq t)]$ of the second term of $\Phi_{ij}$ and using that $(1-e_{j}(t))^2 = (1+e_{j}(t)^2-2e_{j}(t))$. We also re-write the martingale to a counting process and a compensator part, thus we use again $M_{ij}(t)=N_{ij}(t)-\int_{0}^{t}\lambda_{ij}(s)ds$. We end with the following expression for the influence function which is the final result: 
\begin{eqnarray*}
\Phi_{ij}&=& [A_{i}-e_{j}(T_{ij})]I(T_{ij} \leq C_{i}) -\int_{0}^{\tau} [A_{i}-e_{j}(t)] Y_{ij}(t)\gamma_{0j}(t) dt \\
&&-\int_{0}^{\tau}Y_{ij}(t) A_i [1-e_{j}(t)][\gamma_{1j}(t) - \gamma_{0j}(t)]dt \\
&&+\int_{0}^{\tau}[A_{i}Y_{ij}(t)-P(T_{ij}\geq t,\: C_{i}\geq t,\: A_{i}=1)][1-e_{j}(t)]^{2}[\gamma_{1j}(t)-\gamma_{0j}(t)]dt\\
&&+\int_{0}^{\tau}[(1-A_{i})Y_{ij}(t)-P(T_{ij}\geq t,\: C_{i}\geq t,\: A_{i}=0)][e_{j}(t)]^{2}[\gamma_{1j}(t)-\gamma_{0j}(t)]dt.
\end{eqnarray*}

\bibliography{postdocbib}

\end{document}